\documentclass[showpacs,pre,a4paper,preprint]{revtex4}

\usepackage[latin1]{inputenc}
\usepackage{graphicx}

\newcommand{\gcpf}{\ensuremath{\mathcal{Z}}}

\begin{document}

\title{Bethe approximation for a DNA-like self-avoiding walk model with variable solvent quality.}

\author{D P  Foster}
\author{C Pinettes}

\affiliation{Laboratoire de Physique Th\'eorique et Mod\'elisation
(CNRS UMR 8089), Universit\'e de Cergy-Pontoise, 2 ave A. Chauvin
95302 Cergy-Pontoise cedex, France}

\begin{abstract}
The phase diagram and critical behaviour of a simple toy model for DNA zipping/unzipping is examined in the framework of the Bethe approximation. The effects of solvent quality are included, and found to lead to a variety of different thermodynamic behaviours.
\end{abstract}


\maketitle

\section{Introduction}

Self-avoiding walk models and their variants have been used for decades to gain insight into the physics of real polymer systems, with remarkable success\cite{Vanderzande:1998p4444}. Such models are now again of interest as toy models of biopolymers such as DNA and RNA molecules\cite{Baiesi:2003p3677,Leoni:2003p4447,Zara:2007p3673,Zara:2006p3674,Pretti:2006p4448}. The effect of the base pairing is modelled by allowing the walk to visit the lattice bonds twice. In the context of the work presented here, the difference between DNA-like models and  RNA-like models are the allowed walk configurations; DNA models allow at most four links of the walk to meet at a site (see figure~\ref{vertex}), whilst RNA allow up to eight.

Recently several models of this type have been presented, either as models for biopolymers, or simply as interesting variants of restricted random walk models, in which different weightings are given to multiply visited sites or bonds\cite{Oliveira:2008p3601,Serra:2007p4467,Krawczyk:2006p3006}.  What is missing from all these models are interactions representing the interactions with solvent molecules. In this article we propose a first step at filling this void.
 
In this article, we introduce a variant of the lattice two-tolerant self-avoiding walk model\cite{Orlandini:1992p4525,Baiesi:2003p3677,Leoni:2003p4447} which mimics the zipping/unzipping of a DNA model. In the model introduced here we also include solvent effects by including an attractive interaction energy between non-consecutively visited nearest-neighbour sites. In this article we present an extended mean-field type calculation of the phase diagram (the Bethe approximation). The phase diagram is found to be unexpectedly rich.
In many DNA models there is a further restriction in that bases are only allowed to pair up if they are the same distance from one end along each of the two chains. This restriction is relaxed in the model presented here.

\section{Model}

\begin{figure}[h!t]
\begin{center}
\includegraphics[width=12cm,clip]{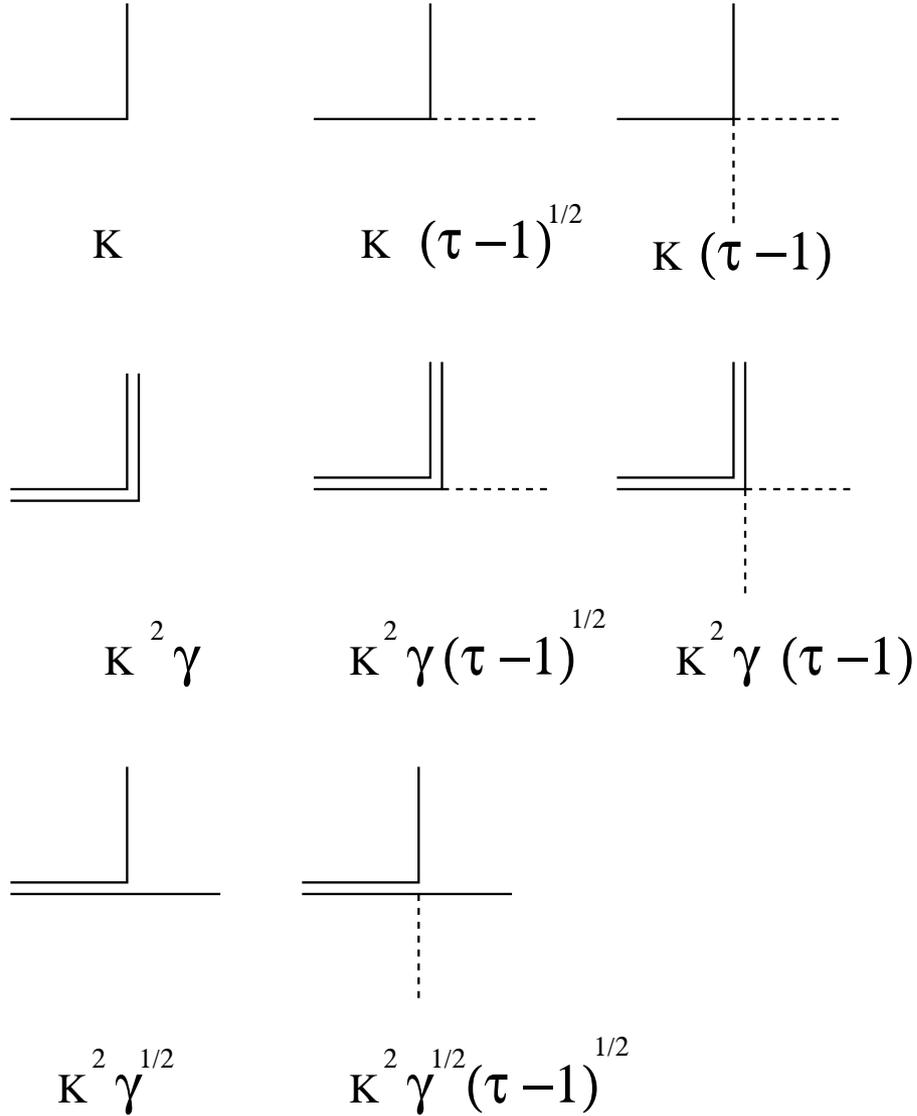}
\end{center}
\caption{Configurations at a lattice site on the square lattice. Polymer bonds are represented by solid lines while the solvent interactions are represented by dashed lines. The empty-site configuration is not shown here.}\label{vertex}
\end{figure}

The model studied in this article consists of a non-crossing random walk on a square lattice limited to visit each bond and each site at most twice. The model is chosen to model the zipping/unzipping of a DNA molecule, thus the allowed configurations of the two-tolerant walk are further restricted such that a site may only be visited twice if one of the adjoining bonds is doubly visited. The allowed configurations are shown in figure~\ref{vertex}.

Each segment of the walk represents a monomer (base) whereas doubly-visited bonds represent a coarse-grained description of paired bases. The difference in affinity of the DNA molecule with itself 
and with the solvent may be modelled by a solvent-mediated attractive interaction between non-consecutive visited nearest-neighbour sites on the lattice.
Solvent-mediated interactions carry an attractive energy $-\varepsilon_{\rm S}$ 
and doubly-visited bonds yield an attractive energy $-\varepsilon$. 

The thermodynamic behaviour may be investigated by introducing the grand-canonical partition function, \gcpf, from which many of the 
relevant thermodynamic quantities may be calculated. The grand-canonical partition function is given by:
\begin{equation}
\gcpf=\sum_{\rm walks} K^N\ \tau^{N_I}\ \gamma^{N_2}
\end{equation}
where $N_I$ is the number of solvent-mediated interactions, $N_2$ is the number of doubly visited bonds and $\tau=\exp\left(-\beta\varepsilon_{\rm S}\right)$, $\gamma=\exp\left(-\beta\varepsilon\right)$ and $\beta=1/kT$. The fugacity, which controls the average length of the walk, is denoted by $K$, and $N$ is the total length of the walk. The two are related through: 
 \begin{equation}
 \langle N\rangle=K\frac{\partial \ln\gcpf}{\partial K}.
 \end{equation}
 The average length increases as $K$ is increased.

\section{The Bethe approximation}\label{bethe}

\begin{figure}[h!t]
\begin{center}
\includegraphics[width=12cm]{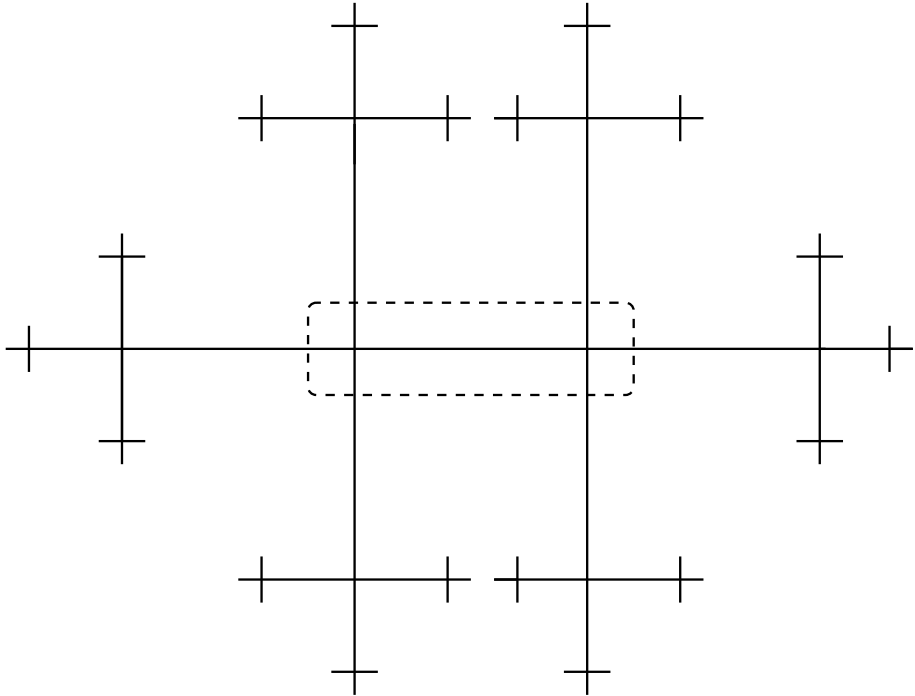}
\end{center}
\caption{The Bethe lattice representation of the  two-dimensional lattice. The dotted box shows the central bond, exhibiting the desired square-lattice geometry. }\label{blat}
\label{Bethe}
\end{figure}

In this section we briefly describe the Bethe approximation. For a good discussion of the Bethe approximaton see Ref.\cite{Baxter:1982gf}.
 The model of interest 
 is studied on the infinite Bethe lattice chosen to have the correct local geometry. The lattice 
 chosen for the square lattice is shown in figure~\ref{blat}. 
 The Bethe lattice is a hierarchical lattice built recursively from a central bond by adding $k$ new bonds to each extremity. To each
dangling bond we add $k$ more bonds, and so on, such that no loops are formed.  Due to the hierarchical nature of the lattice, it is possible to build up expressions for the partition function recursively. To see this, it is convenient to consider the lattice as being divided into two branches,
left and right for the example shown in figure~\ref{blat}. We may introduce the partial partition functions
$W^{\rm l}_\sigma$ and $W^{\rm r}_\sigma$ for the left and right-hand branch, respectively. These 
partition functions are conditional upon the state $\sigma$ of the central bond. 
In our model there are four possible states: 
\begin{enumerate}
\item empty (state 0),
\item occupied with a link of the walk (state $1$),
\item occupied with a solvent-mediated nearest-neighbour interaction (state ${\rm S}$),
\item occupied with a doubly-visited bond (state $2$).
\end{enumerate}

By symmetry, the left and right branches will have the same partial partition functions, and so the l,r designation will be dropped. Each branch may be sub-divided into $k$ sub-branches, such that the $W_\sigma$ may be expressed in terms of the partial partition functions of the sub-branches. 
This procedure may be continued until the boundary bonds are reached. 
In order to do this explicitly, it is convenient to introduce the notion of the `generation' of a link, $n$,
 which is simply the distance of the link from the boundary. As a concrete example, consider the calculation of $W_1^{(n)}$, the partial partition function conditional on the central bond being occupied
 by a link of the walk.
 We must consider all the configurations on the bonds of the generation $(n-1)$, of which there are three for the 2 dimensional square lattice example shown in figure~\ref{blat}, which are compatible with the occupied central bond. Clearly there must be a bond leaving in one of the three directions, the other two bonds may be empty or occupied by a solvent-mediated interaction.  The weight $W_1^{(n)}$ is simply the sum of the Boltzmann weights corresponding to all these configurations, multiplied by the weight for adding the central link. To avoid the divergence of the
 partial partition functions it is 
 convenient to introduce normalised partition functions 
 $w^{(n)}_\sigma=W^{(n)}_\sigma/q_n$\cite{Pretti:2006p4448} with $q_n$ chosen such that:
\begin{equation}
\sum_\sigma w^{(n)}_\sigma=1.
\end{equation}

This leads to recursion relations for the (normalised) partial partition functions:
\begin{equation}
w^{(n)}_\sigma=\frac{\lambda_\sigma}{q_n}\sum_{\{\sigma_i\}} C_{\sigma,\{\sigma_i\}}\prod_{i=1}^k w^{(n-1)}_{\sigma_i},
\end{equation}
where $\{\sigma_i\}$ is the set of states of the $k$ links forming generation $n-1$,  $\lambda_\sigma$ is the Boltzmann weight of the bond added at generation $n$, and the factor $C_{\sigma,\{\sigma_i\}}=1$ 
if the choice of the states $\{\sigma_i\}$ is compatible with the central state $\sigma$, and zero otherwise. 

It is known that there is no phase transition on the infinite Bethe lattice, since the number 
of boundary sites 
grows too rapidly. 
However the recursion relations may be used in the centre of the lattice as self-consistency equations for the two point mean-field theory for the corresponding square lattice. In this case, we assume we have translational invariance, and drop the generational superscripts.
The equilibrium states are then given by solutions of the following set of recursion relations:

\begin{eqnarray}\label{rec1}
w_0&=&\frac{1}{q} \left\{w_0^3+3 w_1^2\left(w_0+w_{\rm S}+w_2\right)+3 w_2^2\left(w_0+w_{\rm S}\right)\right\}\\\label{rec2}
w_1&=&3\frac{K}{q} w_1\left(w_0+w_{\rm S}\right)\left(w_0+w_{\rm S}+2w_2\right)\\\label{rec3}
w_{\rm S}&=&3\frac{(\tau-1)}{q} \left(w_1^2\left(w_0+w_{\rm S}+w_2\right)+w_2^2\left(w_0+w_{\rm S}\right)\right)\\\label{rec4}
w_2&=&3\frac{K^2\gamma}{q} \left(w_0+w_{\rm S}\right)\left(w_1^2+w_2\left(w_0+w_{\rm S}\right)\right)\\\label{rec5}\nonumber
q&=&w_0^3+3w_1^2\left(w_0+w_{\rm S}+w_2\right)+3w_2^2\left(w_0+w_{\rm S}\right)\\
&&+3Kw_1\left(w_0+w_{\rm S}\right)\left(w_0+w_{\rm S}+2w_2\right) \\\nonumber
&&+3K^2\gamma\left(w_0+w_{\rm S}\right)\left(w_1^2+w_2\left(w_0+w_{\rm S}\right)\right)\\\nonumber
&&+3\left(\tau-1\right)\left(w_1^2\left(w_0+w_{\rm S}+w_2\right)+w_2^2\left(w_0+w_{\rm S}\right)\right).
\end{eqnarray}

The partial partition functions give the contribution to one branch of the total partition function, the total (normalised) partition function conditioned upon the state of the central bond 
is then given by the product of the weights for the left and right branches.
Each of the partial partition functions includes the Boltzmann weight corresponding to the state of the central bond, which is thus counted twice in the full partition function. This double counting is 
corrected by dividing each term by the relevant Boltzmann weight. 
Summing over all the possible states for the central bond gives the total (normalised) partition function, $z$:
\begin{equation}
z=\sum_\sigma \frac{w_\sigma^2}{\lambda_\sigma}.
\end{equation}
In the usual way, the probability of finding a given bond in state $\sigma$ is given by the partition function
conditioned upon this state divided by the total partition function, i.e.
\begin{equation}
p_\sigma=\frac{w^2_\sigma}{z\lambda_\sigma}.
\end{equation}
It should be noted that the density $\rho$ of the walk on the lattice is simply 
\begin{equation}
\rho=p_1+2p_2
\end{equation}
Another quantity of interest is the fraction of paired segments, given by 
\begin{equation}
\Phi=\frac{2p_2}{\rho}
\end{equation}

The grand potential per site may be related to $z$ and $q$ through the relation
\begin{equation}
\beta f=\frac{(k-1)\ln z -2\ln q}{2 } =\ln z-\ln q,
\end{equation}
for the square lattice ($k=3$). For a full derivation of this expression see~\cite{Pretti:2006p4448}.
When multiple solutions to the recurrence relations exist, 
the solution with the lowest value of the grand potential is the stable equilibrium solution. 

\section{Results}\label{results}

It is convenient to recast the recursion relations ~\ref{rec1}---\ref{rec4}  by setting:
\begin{eqnarray*}
x_1&=&\frac{w_1}{w_0},\\
x_{\rm S}&=&\frac{w_{\rm S}}{w_0},\\
x_2&=&\frac{w_2}{w_0}.
\end{eqnarray*} 

\begin{figure}[h!t]
\begin{center}
\includegraphics[width=12cm,clip]{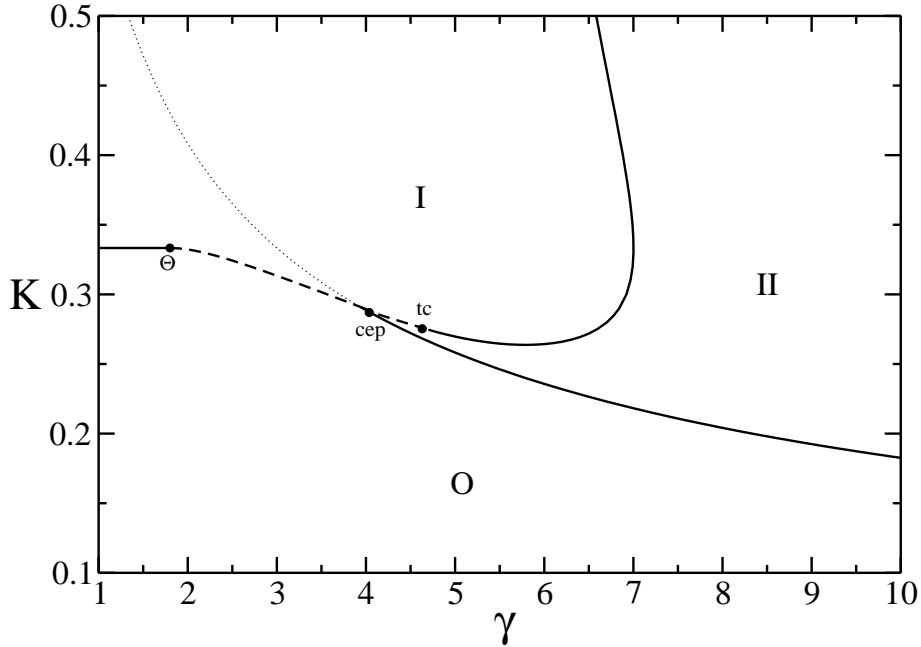}
\end{center}
\caption{The phase diagram in the $K-\gamma$ plane, with $\tau=1$. O denotes the zero density phase, I the ordinary dense phase and II the fully paired dense phase. Dashed and solid lines denote first- and second-order transitions respectively. The tricritical $\Theta$ point on the O-I transition line is given by $K=1/3$ and $\gamma_\Theta=9/5$. The second-order transition between phases O and II is given by the line $3K^2\gamma=1$, and finishes in a critical end point. The dotted line is the extension of this line and is included to guide the eye.}\label{pd1}
\end{figure}

The different phases and transition lines correspond to different solutions of the recursion equations.
These equations have a trivial solution $x_1=x_{\rm S}=x_2=0$, corresponding to the finite polymer phase. The density is trivially zero, since the lattice is infinite. In what follows we refer to this phase as the  O phase. Non-trivial solutions may be found  by setting different values of the parameters $K$, $\tau$ and $x_{\rm S}$ and then solving numerically for $\gamma$, $x_1$ and $x_2$.

The trivial zero-density phase is separated from a region where the walk fills the lattice with a finite density by a surface $K=K_\infty(\tau,\gamma)$ where the average length of the walk first diverges.
In the finite-density region we find
two dense phases :
a phase (I) in which a finite fraction of the segments of the walk are paired, i.e. it is characterised 
by $0<\rho<1$ and $0<\Phi<1$ ;
and a fully paired dense phase  (II) in which every segment is paired i.e. it is characterized by $0<\rho<1$ and $\Phi=1$.


When $\gamma=0$ (giving $x_2\equiv 0$) the pairing of segments is forbidden, and the model corresponds to the standard interacting self-avoiding walk model. For this model, also called the $\Theta$-point model\cite{Vanderzande:1998p4444}, phases O and I are separated by a critical transition line for small enough $\tau$ and a first order line for large enough  $\tau$. These two behaviours are separated by the tricritical $\Theta$-point. This tricritical point extends to a line of tricritical points as $\gamma$ is allowed to increase, separating a critical region in the SAW universality class from the first order region, where  the walk fills the lattice with a finite density.
It is possible to determine analytically the region occupied by the SAW phase, and so the equation of the tricritical line, as follows: from phase I ($x_1\ne0$) we take the limit $x_{\rm S}\rightarrow 0$ and $x_2\rightarrow 0$ in order to approach the boundary with phase O. 
Since the transition is continuous, this limit leads to $x_1\rightarrow0$ and gives $K_\infty=1/(2d-1)=1/3$.
The tricritical point is obtained by allowing the parameters $x_1$, $x_{\rm S}$ and $x_2$ to be vanishingly small but non-zero along the transition line $K=1/3$. We obtain the tricritical condition :
\begin{equation}
\gamma_{\rm \Theta}(\tau)=\frac{9(3-2\tau)}{4(\tau-1)+5(3-2\tau)}.
\end{equation}
It is important to note, that there is no guarantee that the extended line is in the same universality class as the $\Theta$-point, but we will nevertheless use the $\Theta$ subscript to differentiate it from other special points in the phase diagram. We shall return to this point later.
If the pairing interaction is taken to be attractive, then $\gamma\geq 1$, this leads to the result that the transition line O-I becomes fully first order for $\tau>4/3$.

\begin{figure}[h!t]
\begin{center}
\includegraphics[width=12cm,clip]{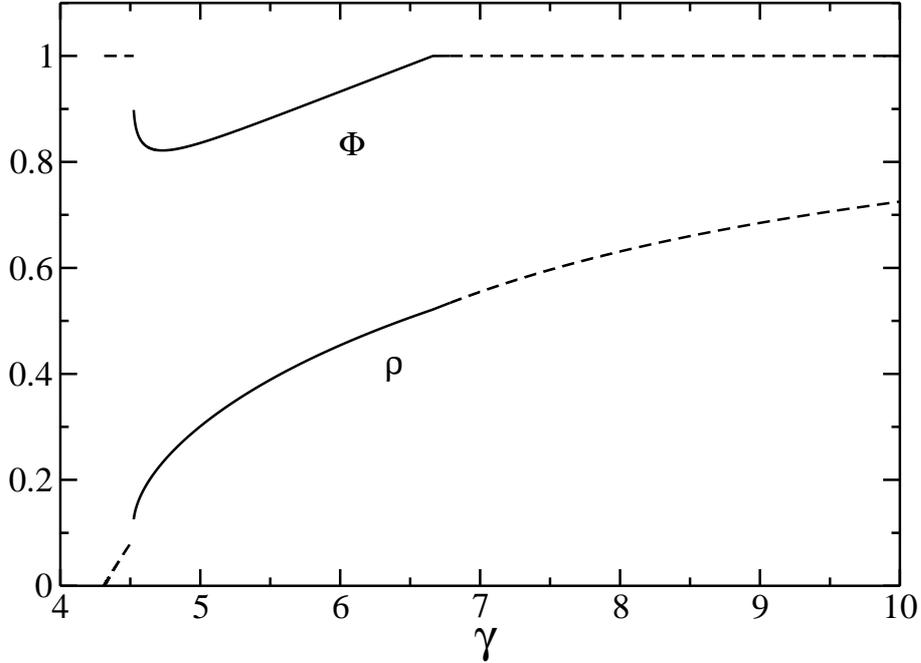}
\end{center}
\caption{The two order parameters, the density $\rho$ and the fraction of paired segments $\Phi$ as a function of $\gamma$ for $\tau=1$ and $K=0.278$. Solid and dashed lines denote their values in phases  I and II.}\label{op1}
\end{figure}


\begin{figure}[h!t]
\begin{center}
\includegraphics[width=12cm,clip]{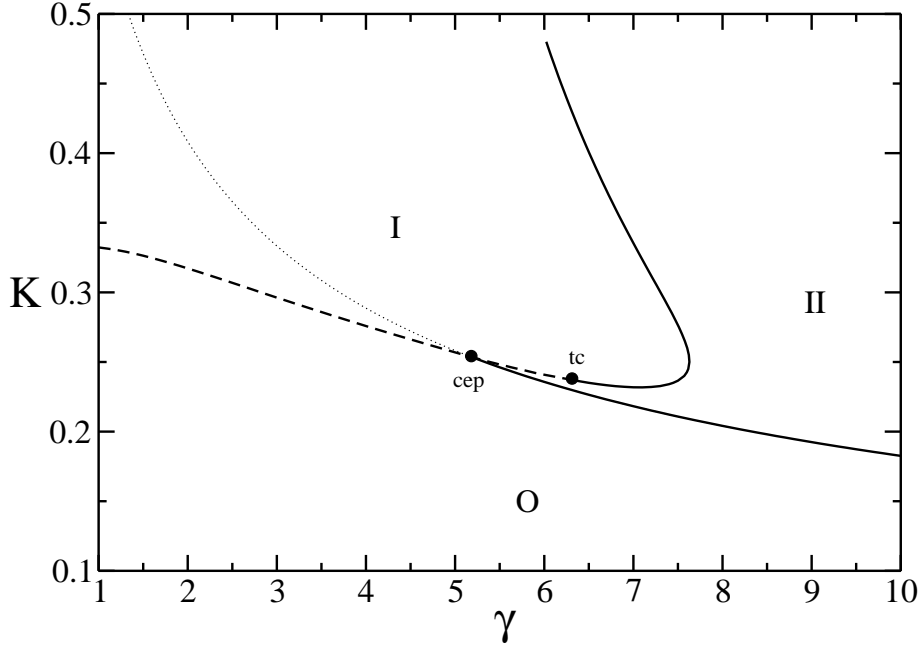}
\end{center}
\caption{The phase diagram in the $K-\gamma$ plane, with $4/3<\tau=1.4<3/2$. Dashed and solid lines denote first- and second-order transitions respectively. The point marked cep is a critical end point, whilst tc indicates a tricritical point. The dotted line corresponds to the line $3K^2\gamma=1$.}\label{pd2}
\end{figure}

\begin{figure}[h!t]
\begin{center}
\includegraphics[width=12cm,clip]{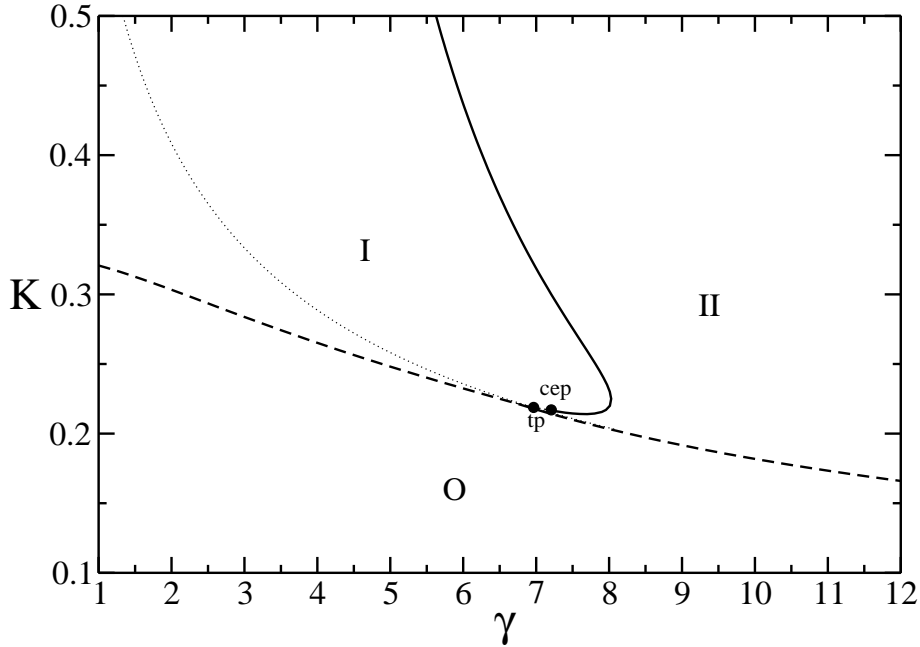}
\end{center}
\caption{The phase diagram in the $K-\gamma$ plane, with $3/2<\tau=1.6<1.75$. Dashed and solid lines denote first- and second-order transitions respectively. The point marked tp is a triple point, whilst cep denotes a critical end point. The dotted line corresponds to the line $3K^2\gamma=1$.}\label{pd3}
\end{figure}

\begin{figure}[h!t]
\begin{center}
\includegraphics[width=12cm,clip]{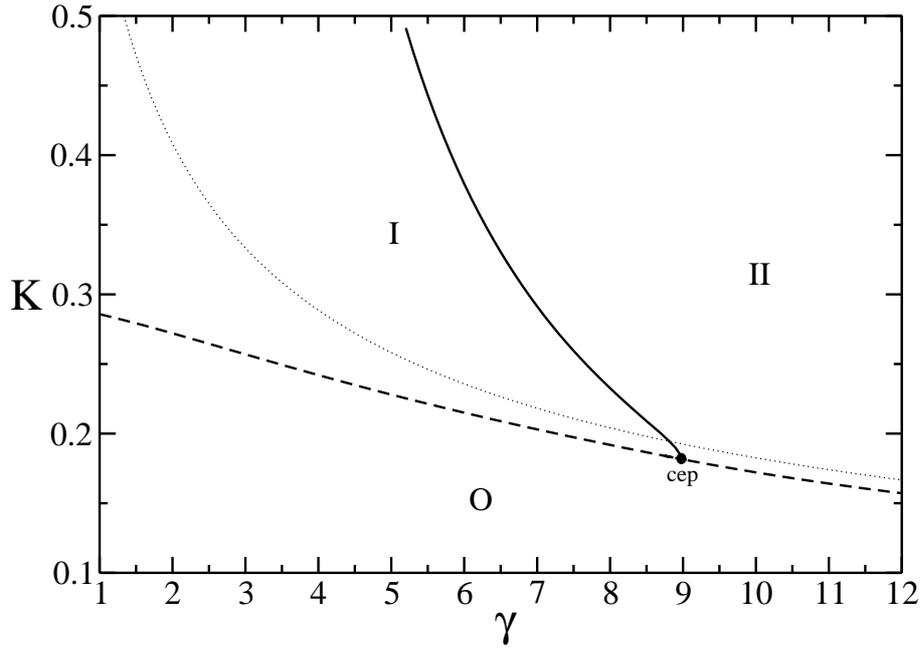}
\end{center}
\caption{The phase diagram in the $K-\gamma$ plane, with $\tau=2>1.75$. Dashed and solid lines denote first- and second-order transitions respectively. The dotted line corresponds to the line $3K^2\gamma=1$.}\label{pd4}
\end{figure}

The phase diagram for $\tau=1$ is shown in figure~\ref{pd1}, clearly showing the location of the tricritical $\Theta$-like point, found to be $K=1/3$ and $\gamma_\Theta=9/5$.
The phase boundary of phase I is smooth, with the boundary between phases I and II being partly first order and partly second order. This
gives rise
to another tricritical point at $\gamma_{\rm tc}\approx 4.7$. The transition line between phases O and II is found to be continuous, and the equation for this line may be determined exactly: 
from phase II ($x_1=0$ and $x_2\neq0$) we take the limit $x_{\rm S}\rightarrow0$ to approach the boundary with phase O. Using the fact that the transition is continuous, we determine that $x_2\rightarrow0$ leads to the condition $3K^2\gamma=1$. This line terminates in a critical end point at $\gamma_{\rm cep}\approx 3.94$.

For $\tau=1$, i.e. in absence of solvent effects, the phase diagram is similar to the phase diagram 
obtained by Pretti for an RNA-like model with non-zero stacking energy \cite{Pretti:2006p4448}, except that in Pretti's case the equivalent to the O-II line was found to be first order. The stacking energy had the effect of favouring the absence of multiple double bonds meeting at a site, and so made the model more like the model presented here.



\begin{figure}[h!t]
\begin{center}
\includegraphics[width=12cm,clip]{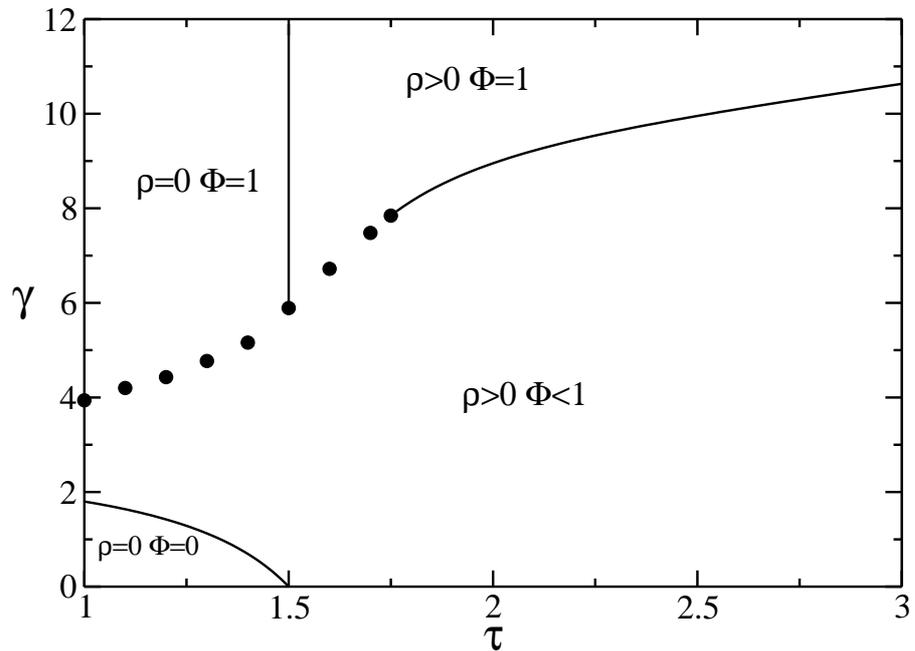}
\end{center}
\caption{Phase diagram in the $K_\infty(\gamma,\tau)$ plane. Solid lines represent continuous phase transitions, whilst the points indicate a first-order transition line.}\label{pdkinf}
\end{figure}

\begin{figure}[h!t]
\begin{center}
\includegraphics[width=12cm,clip]{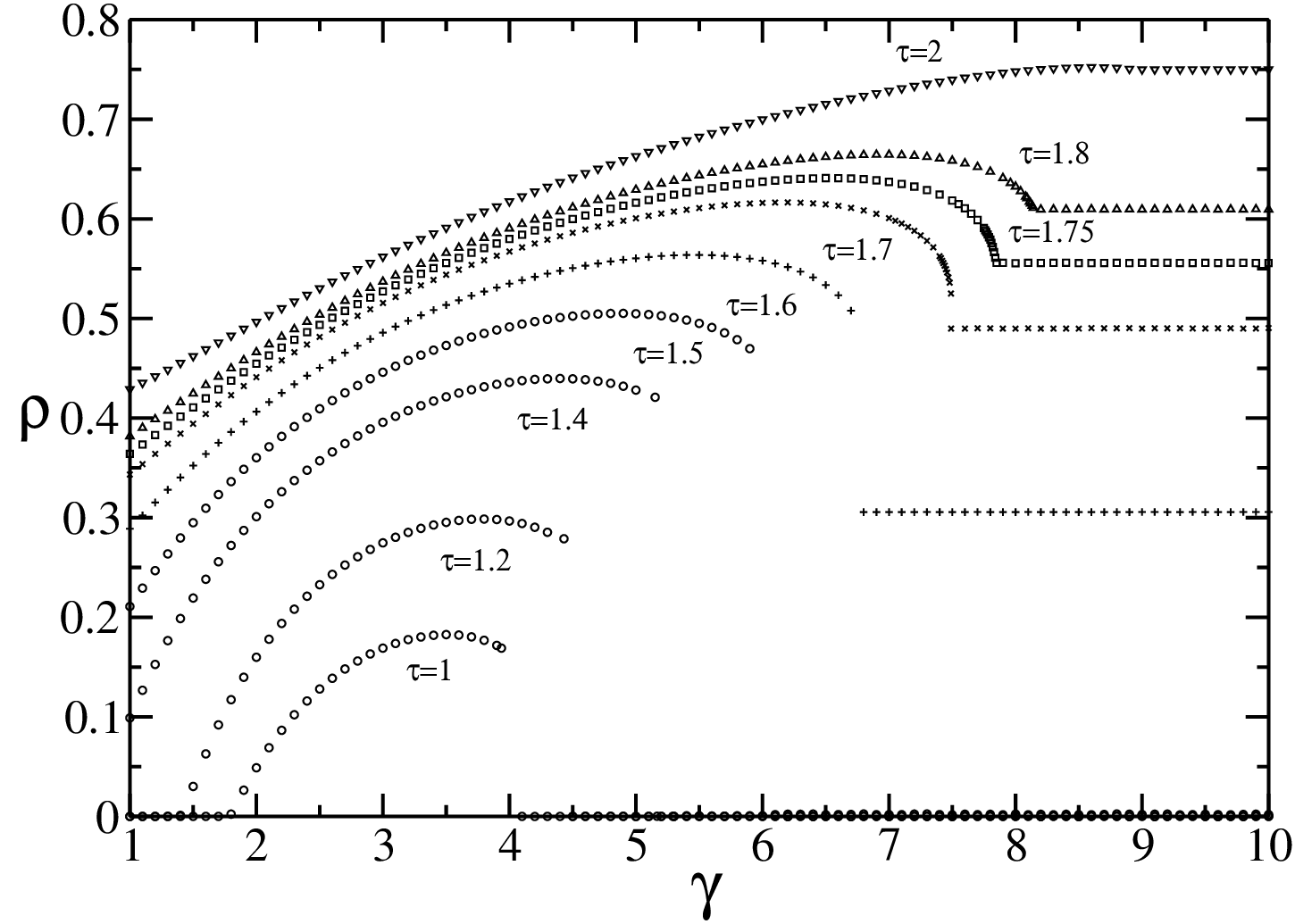}
\end{center}
\caption{Density $\rho$ plotted in the  $K_\infty (\gamma,\tau)$ plane as a function of $\gamma$ for different values of $\tau$.}\label{first}
\end{figure}

Before examining other values of $\tau$, it is of interest to discuss the different phases shown in figure~\ref{pd1}, and the relevant order parameters. Two parameters are of interest for differentiating the different phases: the density of occupied lattice bonds $\rho$, and the fraction of paired segments $\Phi$. These two quantities are plotted in figure~\ref{op1} for $K=0.278$, so as to pass through each phase.
In phase O $\rho=0$ trivially since the walk is finite on an infinite lattice. In the framework of the Bethe approximation it is not possible to calculate $\Phi$ in this phase ($\rho=0, p_2=0$). The value of $K$ is chosen to cross the transition from phase O to phase II under the critical end point. We clearly see that $\rho$ increases smoothly from $0$, indicative of a critical transition, and that $\Phi=1$ indicating that phase II corresponds to a saturated doubly occupied phase. Again we remind the reader that the apparent jump in $\Phi$ is simply related to the fact that it is not possible to calculate $\Phi$ in phase O (where $p_1=p_2=0$).
The value of K is chosen to cut back into phase I between the critical end point and the second tricritical point. We can clearly see in figure~\ref{op1} that $\Phi$ drops discontinuously from $1$ and there is a small jump in $\rho$. The transition from phase I to phase II is again continuous, and the interest of $\Phi$ as order parameter is clearly seen; there is no evidence of a phase transition in $\rho$, which is not a good order parameter for phase II. 

\begin{figure}[h!t]
\begin{center}
\includegraphics[width=12cm,clip]{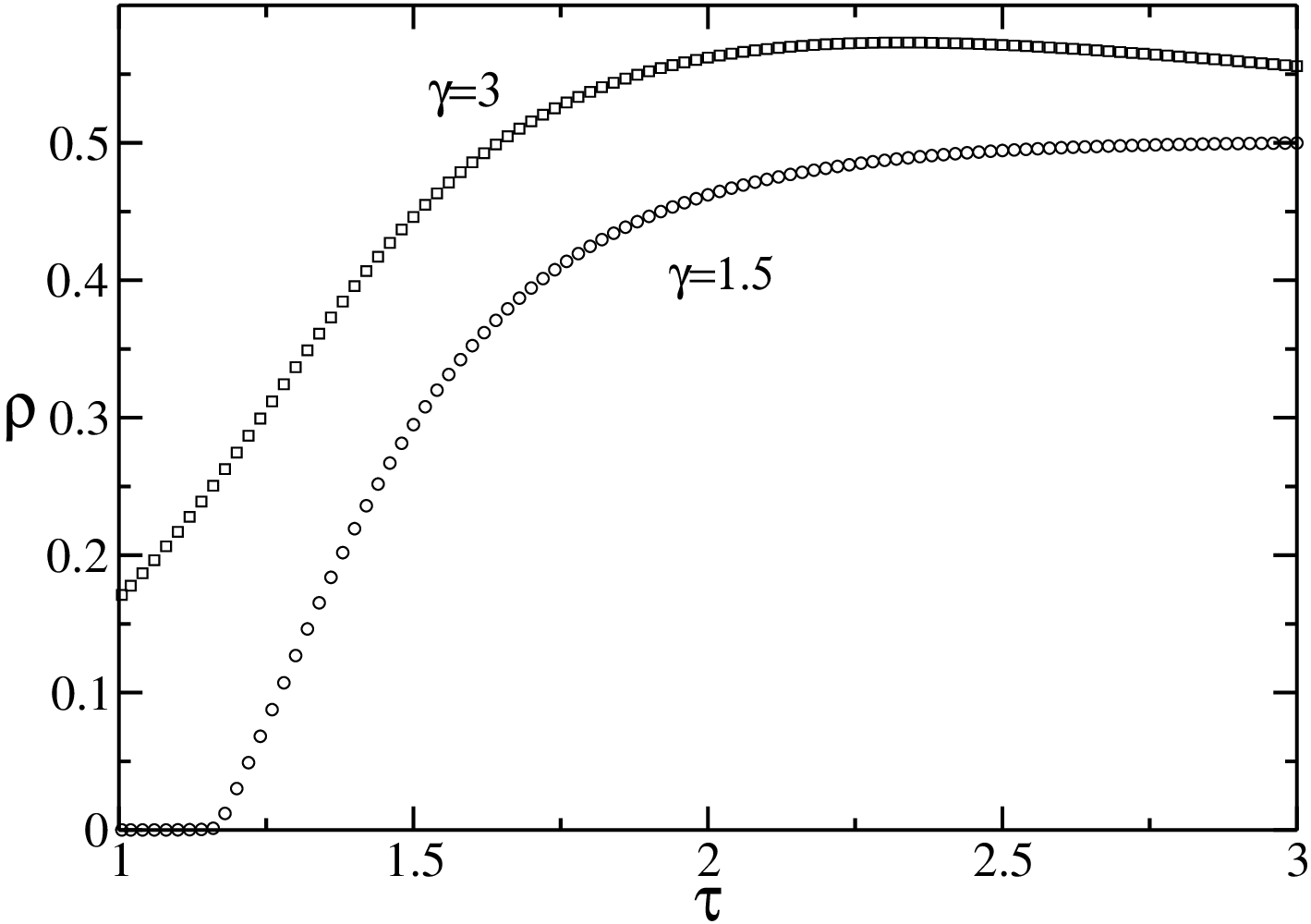}
\end{center}
\caption{Density $\rho$ plotted in the  $K_\infty (\gamma,\tau)$ plane as a function of $\tau$ for different values of $\gamma$.}
\end{figure}

\begin{figure}[h!t]
\begin{center}
\includegraphics[width=12cm,clip]{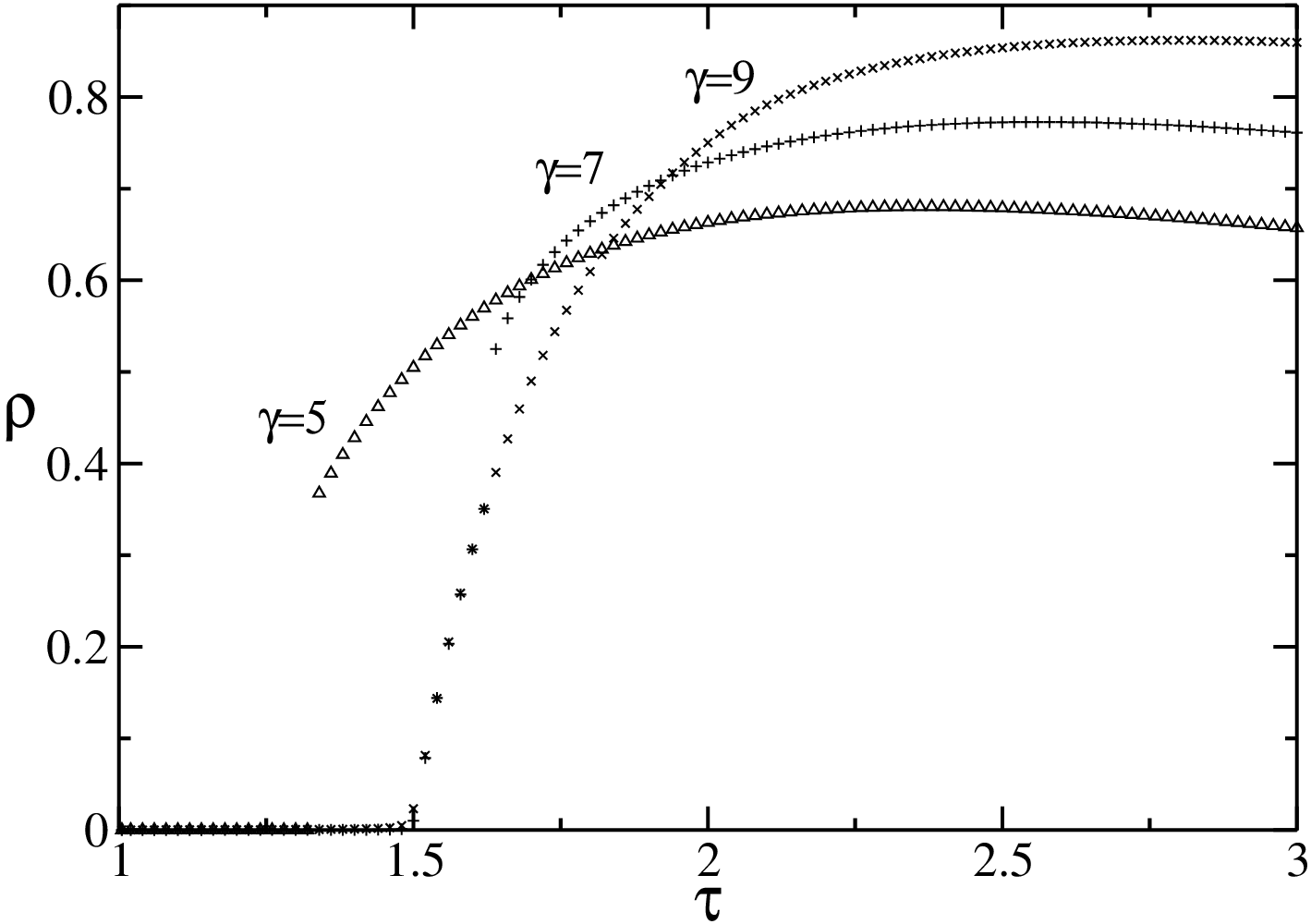}
\end{center}
\caption{Density $\rho$ plotted in the  $K_\infty (\gamma,\tau)$ plane as a function of $\tau$ for different values of $\gamma$.}
\end{figure}

As $\tau$ is increased, the phase diagram changes. The first change is that the $\Theta$-like tricritical transition is pushed out of the domain of interest ($\varepsilon>0$ or $\gamma>1$). This occurs, as previously stated, for $\tau=4/3$. This case is shown in figure~\ref{pd2}.
The calculation for the transition line between phases O and II remains valid as $\tau$ is increased, but for large enough $\tau$ the transition is found to be first order, see the phase diagrams in figures~\ref{pd3} and~\ref{pd4}. We may determine the condition for this change in behaviour by allowing the parameters $x_{\rm S}$ and $x_2$ to be vanishingly small but non-zero along the transition line. We obtain the condition $\tau=3/2$ . Thus the O-II line is second order and given by the equation $3K^2\gamma=1$ for  $\tau\leq3/2$ and it is first order for $\tau>3/2$, and falls below the line $3K^2\gamma=1$. This behaviour may be understood in another way; as $\gamma\to\infty$, the walk becomes a doubled up self-avoiding walk, with an effective fugacity per bond $\tilde{K}=K^2\gamma$. We may then apply the known results to find that for $\tilde{K}<1/3$ the transition from phase O to the dense phase is continuous, and for $\tilde{K}>1/3$ the transition is first-order. There is a $\Theta$ transition when $\tau=3/2$ and $3\tilde{K}=3K^2\gamma=1$. Since phase II is a saturated phase, the walk remains a doubled up self-avoiding walk for finite $\gamma$, hence the observed result, and the prediction that for $\tau=3/2$ the transition line between phase O and phase II is of the same type as the $\Theta$ transition. It would be interesting to ascertain whether this prediction is maintained if a full calculation for the model is performed. For $\tau<3/2$ the first-order line between phases O and I and the first-order line between phases I and II are tangential to each other at the critical end point which defines the end of the critical line between phases O and II. However, for $3/2<\tau<1.75$ (see figure~\ref{pd3}) the  three first-order lines meet at a triple point (however the first-order lines between phases O and I  and between I and II are still tangential).

Another interesting change arises when $\tau\approx 1.75$ (see figure~\ref{pd4}). The phase transition between phases I and II becomes  totally second-order, and the first-order boundary to phase O becomes smooth, and the point where the three phases meet is again a critical end point.

\begin{figure}[h!t]
\begin{center}
\includegraphics[width=12cm,clip]{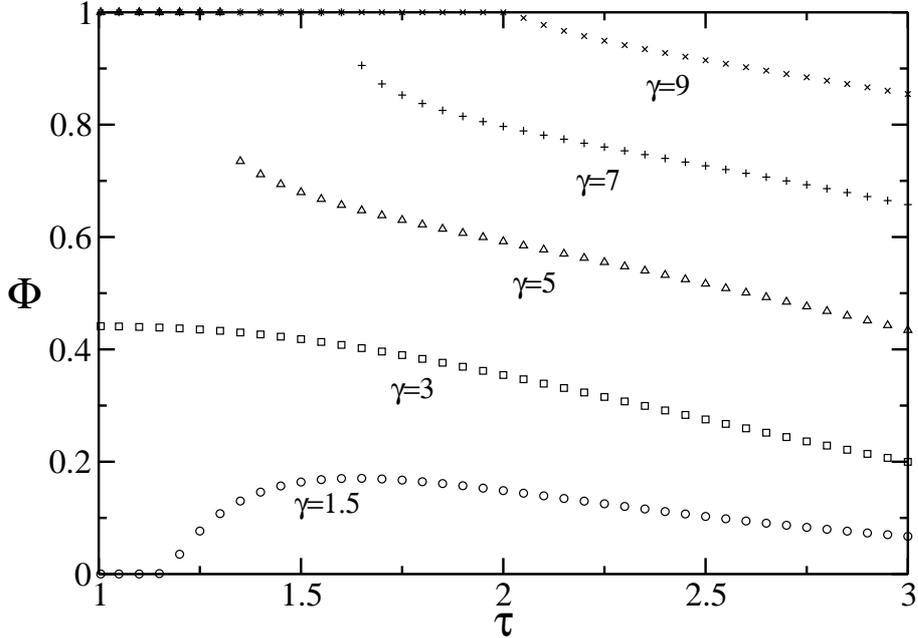}
\end{center}
\caption{Fraction of paired bonds $\Phi$ plotted in the  $K_\infty (\gamma,\tau)$ plane as a function of $\tau$ for different values of $\gamma$.}\label{last}
\end{figure}

Of particular interest is the phase diagram in the plane $K_\infty(\gamma,\tau)$, where the average length of the walk first diverges. This plane is particularly important, since it corresponds to the phase diagram which is seen in a canonical simulation of single finite-sized walk in the limit of infinite walk length. Indeed it is sometimes referred to as the ``thermodynamic limit" for the canonical dilute polymer problem. The phase diagram in the $K_\infty$ plane is shown in figure~\ref{pdkinf}. There are four different phases, differentiated by the values of $\rho$ and $\Phi$. There are two critical $\rho=0$ phases, both of which should be in the self-avoiding walk universality class. One has $\Phi=0$ (the standard SAW phase) whilst the other has $\Phi=1$ corresponding to a double SAW phase. There are two non-critical phases, one with $\Phi=1$, corresponding to a standard type of collapsed phase as seen in the $\Theta$-point model, except that the walk is doubled. The other phase has a value of $\Phi$ different from both $0$ and $1$. 

It is interesting to note that $\gamma_\Theta(\tau)$ decreases as $\tau$ increases, making it easier for the walk to pass from the $\rho=0,\Phi=0$ phase to the neighbouring dense phase as $\tau$ is increased. At first sight this seems odd, however the formation of some double bonds enables the formation of a branched polymer type conformation, which is more naturally space filling. This would lead one to expect that perhaps the phase boundary between the $\rho=0, \Phi=0$ phase and the $\rho>0, 0<\Phi<1$ phase may not be in the $\Theta$-point universality class. The mean-field nature of the Bethe approximation does not enable one to investigate this point. 

From previous discussion, the line separating the two $\Phi=1$ phases, found at $\tau=3/2$, may be expected to be a line of tricritical points in the $\Theta$-point universality class. This line appears to end at a tricritical end point on a first-order line ($\tau=3/2, \gamma\approx 5.89$), which in its turn  ends at another tricritical point  at $\tau\approx 1.75, \gamma\approx 7.845$. Whilst much could change once fluctuations are included, one could expect much of the behaviour seen here to remain in the real system. Figures~\ref{first} to~\ref{last} show different plots of $\rho$ and $\Phi$ showing their behaviour at the different phase boundaries, confirming the orders of the transitions shown.

\section{Discussion}\label{discussion}

In this article we have investigated the phase diagram of a non-crossing random walk model on the square lattice where the walk is allowed to visit lattice bonds twice, and the site configurations are chosen to represent the zipping/unzipping of a DNA molecule. Solvent quality is included through attractive nearest-neighbour interactions. This model is examined in the framework of the Bethe approximation and found to have a rich phase diagram. Whilst there are limitations to the method, it usually captures the essential features of the phase diagram\cite{Foster:2008p4129,Lise:1998fj}. It would be of interest to  look at this model using some other method to confirm and examine further the phase diagram presented.


\begin{thebibliography}{12}
\expandafter\ifx\csname natexlab\endcsname\relax\def\natexlab#1{#1}\fi
\expandafter\ifx\csname bibnamefont\endcsname\relax
  \def\bibnamefont#1{#1}\fi
\expandafter\ifx\csname bibfnamefont\endcsname\relax
  \def\bibfnamefont#1{#1}\fi
\expandafter\ifx\csname citenamefont\endcsname\relax
  \def\citenamefont#1{#1}\fi
\expandafter\ifx\csname url\endcsname\relax
  \def\url#1{\texttt{#1}}\fi
\expandafter\ifx\csname urlprefix\endcsname\relax\def\urlprefix{URL }\fi
\providecommand{\bibinfo}[2]{#2}
\providecommand{\eprint}[2][]{\url{#2}}

\bibitem[{\citenamefont{Vanderzande}(1998)}]{Vanderzande:1998p4444}
\bibinfo{author}{\bibfnamefont{C.}~\bibnamefont{Vanderzande}},
 {\it Lattice Models of Polymers} (Cambridge University Press, Cambridge, England, \bibinfo{year}{1998}).

\bibitem[{\citenamefont{Baiesi et~al.}(2003)\citenamefont{Baiesi, Orlandini,
  and Stella}}]{Baiesi:2003p3677}
\bibinfo{author}{\bibfnamefont{M.}~\bibnamefont{Baiesi}},
  \bibinfo{author}{\bibfnamefont{E.}~\bibnamefont{Orlandini}},
  \bibnamefont{and} \bibinfo{author}{\bibfnamefont{A.~L.}~\bibnamefont{Stella}},
  \bibinfo{journal}{Phys Rev Lett} \textbf{\bibinfo{volume}{91}},
  \bibinfo{pages}{198102} (\bibinfo{year}{2003}).

\bibitem[{\citenamefont{Leoni and Vanderzande}(2003)}]{Leoni:2003p4447}
\bibinfo{author}{\bibfnamefont{P.}~\bibnamefont{Leoni}} \bibnamefont{and}
  \bibinfo{author}{\bibfnamefont{C.}~\bibnamefont{Vanderzande}},
  \bibinfo{journal}{Physical Review E} \textbf{\bibinfo{volume}{68}},
  \bibinfo{pages}{051904} (\bibinfo{year}{2003}).

\bibitem[{\citenamefont{Pretti}(2006)}]{Pretti:2006p4448}
\bibinfo{author}{\bibfnamefont{M.}~\bibnamefont{Pretti}},
  \bibinfo{journal}{Physical Review E} \textbf{\bibinfo{volume}{74}},
  \bibinfo{pages}{051803} (\bibinfo{year}{2006}).

\bibitem[{\citenamefont{Zara and Pretti}(2006)}]{Zara:2006p3674}
\bibinfo{author}{\bibfnamefont{R.~A.} \bibnamefont{Zara}} \bibnamefont{and}
  \bibinfo{author}{\bibfnamefont{M.}~\bibnamefont{Pretti}},
  \bibinfo{journal}{Physica A} \textbf{\bibinfo{volume}{371}},
  \bibinfo{pages}{88} (\bibinfo{year}{2006}).


\bibitem[{\citenamefont{Zara and Pretti}(2007)}]{Zara:2007p3673}
\bibinfo{author}{\bibfnamefont{R.~A.} \bibnamefont{Zara}} \bibnamefont{and}
  \bibinfo{author}{\bibfnamefont{M.}~\bibnamefont{Pretti}}, \bibinfo{journal}{J
  Chem Phys} \textbf{\bibinfo{volume}{127}}, \bibinfo{pages}{184902}
  (\bibinfo{year}{2007}).


\bibitem[{\citenamefont{Krawczyk et~al.}(2006)\citenamefont{Krawczyk,
  Prellberg, Owczarek, and Rechnitzer}}]{Krawczyk:2006p3006}
\bibinfo{author}{\bibfnamefont{J.}~\bibnamefont{Krawczyk}},
  \bibinfo{author}{\bibfnamefont{T.}~\bibnamefont{Prellberg}},
  \bibinfo{author}{\bibfnamefont{A.~L.} \bibnamefont{Owczarek}},
  \bibnamefont{and}
  \bibinfo{author}{\bibfnamefont{A.}~\bibnamefont{Rechnitzer}},
  \bibinfo{journal}{Phys Rev Lett} \textbf{\bibinfo{volume}{96}},
  \bibinfo{pages}{240603} (\bibinfo{year}{2006}).

\bibitem[{\citenamefont{Serra and Stilck}(2007)}]{Serra:2007p4467}
\bibinfo{author}{\bibfnamefont{P.}~\bibnamefont{Serra}} \bibnamefont{and}
  \bibinfo{author}{\bibfnamefont{J.~F.} \bibnamefont{Stilck}},
  \bibinfo{journal}{Physical Review E} \textbf{\bibinfo{volume}{75}},
  \bibinfo{pages}{011130} (\bibinfo{year}{2007}).

\bibitem[{\citenamefont{Oliveira et~al.}(2008)\citenamefont{Oliveira, Stilck,
  and Serra}}]{Oliveira:2008p3601}
\bibinfo{author}{\bibfnamefont{T.~J.} \bibnamefont{Oliveira}},
  \bibinfo{author}{\bibfnamefont{J.~F.} \bibnamefont{Stilck}},
  \bibnamefont{and} \bibinfo{author}{\bibfnamefont{P.}~\bibnamefont{Serra}},
  \bibinfo{journal}{Physical Review E} \textbf{\bibinfo{volume}{77}},
  \bibinfo{pages}{041103} (\bibinfo{year}{2008}).

\bibitem{Orlandini:1992p4525} 
\bibinfo{author}{\bibfnamefont{E.}~\bibnamefont{Orlandini}},
  \bibinfo{author}{\bibfnamefont{F.}~\bibnamefont{Seno}},
  \bibinfo{author}{\bibfnamefont{A.~L.} \bibnamefont{Stella}},
  \bibnamefont{and}
  \bibinfo{author}{\bibfnamefont{M.~C.}~\bibnamefont{Tesi}},
  \bibinfo{journal}{Phys Rev Lett} \textbf{\bibinfo{volume}{68}},
  \bibinfo{pages}{488} (\bibinfo{year}{1992}).

\bibitem[{\citenamefont{Baxter}(1982)}]{Baxter:1982gf}
\bibinfo{author}{\bibfnamefont{R.}~\bibnamefont{Baxter}},
  \emph{\bibinfo{title}{Exactly Solved Models in Statistical Mechanics}}
  (Academic Press, New York, \bibinfo{year}{1982}).

\bibitem[{\citenamefont{Lise et~al.}(1998)\citenamefont{Lise, Maritan, and
  Pelizzola}}]{Lise:1998fj}
\bibinfo{author}{\bibfnamefont{S.}~\bibnamefont{Lise}},
  \bibinfo{author}{\bibfnamefont{A.}~\bibnamefont{Maritan}}, \bibnamefont{and}
  \bibinfo{author}{\bibfnamefont{A.}~\bibnamefont{Pelizzola}},
  \bibinfo{journal}{Phys. Rev. E} \textbf{\bibinfo{volume}{58}},
  \bibinfo{pages}{R5241} (\bibinfo{year}{1998}).


\bibitem[{\citenamefont{Foster and Aniambossou}(2008)}]{Foster:2008p4129}
\bibinfo{author}{\bibfnamefont{D.~P.} \bibnamefont{Foster}} \bibnamefont{and}
  \bibinfo{author}{\bibfnamefont{M.}~\bibnamefont{Aniambossou}},
  \bibinfo{journal}{Phys. Rev. E} \textbf{\bibinfo{volume}{77}},
  \bibinfo{pages}{061121} (\bibinfo{year}{2008}).

\end{thebibliography}

\end{document}